\documentclass[journal,onecolumn]{IEEEtran}
\IEEEoverridecommandlockouts
\usepackage{cite}
\usepackage{amsmath,amssymb,amsfonts}
\usepackage{algorithmic}
\usepackage{graphicx}
\usepackage{textcomp}
\usepackage{xcolor}
\def\BibTeX{{\rm B\kern-.05em{\sc i\kern-.025em b}\kern-.08em
    T\kern-.1667em\lower.7ex\hbox{E}\kern-.125emX}}

\usepackage{booktabs} 
\usepackage{enumerate}
\usepackage{hyperref}
\usepackage{adjustbox}
\usepackage{footnote}
\usepackage{enumitem}

\usepackage{multirow}
\usepackage{longtable,csquotes}
\usepackage{subcaption}
\usepackage{comment}
\usepackage{arydshln}
\usepackage{dirtytalk}
\usepackage{url}
\usepackage{xcolor}
\usepackage{pifont} 

\usepackage{graphicx}

\usepackage{caption}
\usepackage{subcaption}

\usepackage{algorithm}
\usepackage{algorithmic}

\begin{document}

\title{The Accuracy of Domain Specific and Descriptive Analysis Generated by Large Language Models}

\author{\IEEEauthorblockN{Denish Omondi Otieno\IEEEauthorrefmark{1},
Faranak Abri\IEEEauthorrefmark{2}, Sima Siami-Namini\IEEEauthorrefmark{3} and
Akbar Siami Namin\IEEEauthorrefmark{1}} \\
\IEEEauthorblockA{\IEEEauthorrefmark{1}Texas Tech University, \IEEEauthorrefmark{2}San Jose State University, \IEEEauthorrefmark{3}Johns Hopkins University\\
Email: \IEEEauthorrefmark{1}deotieno@ttu.edu,
\IEEEauthorrefmark{2}faranak.abri@sjsu.edu,
\IEEEauthorrefmark{1}akbar.namin@ttu.edu,
\IEEEauthorrefmark{3}ssiamin1@jhu.edu}}

\maketitle

\begin{abstract}
Large language models (LLMs) have attracted considerable attention as they are capable of showcasing impressive capabilities generating comparable high-quality responses to human inputs. 
LLMs, can not only compose textual scripts such as emails and essays but also executable programming code. 
Contrary, the automated reasoning capability of these LLMs in performing statistically-driven descriptive analysis, particularly on user-specific data and as personal assistants to users with limited background knowledge in an application domain who would like to carry out basic, as well as advanced statistical and domain-specific analysis is not yet fully explored. More importantly, the performance of these LLMs has not been compared and discussed in detail when domain-specific data analysis tasks are needed. Additionally, the use of LLMs in isolation is often at times insufficient for creating powerful applications and the real potential comes when LLMs are combined with other sources of computation such as LangChain.
This study, consequently, explores whether LLMs can be used as generative AI-based personal assistants to users with minimal background knowledge in an application domain infer key data insights. 
To demonstrate the performance of the LLMs, the study reports a case study through which descriptive statistical analysis, as well as Natural Language Processing (NLP) based investigations, are performed on a number of phishing emails with the objective of comparing the accuracy of the results generated by LLMs to the ones produced by analysts.
The experimental results show that LangChain and the Generative Pre-trained Transformer (GPT-4) excel in numerical reasoning tasks i.e., temporal statistical analysis, achieve competitive correlation with human judgments on feature engineering tasks while struggle to some extent on domain specific knowledge reasoning, where domain-specific knowledge is required. 
\end{abstract}

\begin{IEEEkeywords}
Large Language Models, LangChain, Generative Pre-trained Transformer, Natural Language Processing, Phishing Emails.
\end{IEEEkeywords}

\vspace{-0.09in}
\section{Introduction}
\label{sec:intro}

The debut of Generative Pretrained transformer-based technologies such as GPTs has marked a notable progression in Large Language Models (LLMs), showcasing their impressive and exceptional capabilities through training and, in certain instances, with reinforcement learning from human feedback  \cite{li2023chatgpt, christiano2017deep}. LLMs, have been integrated with robust reasoning engines that empower them to deliberate decisions and suggest actions that need to be taken in unseen circumstances which mirror the data provided to them during training. Effectively, LLMs are achieving remarkable outcomes in a wide range of tasks such as multi-turn question answering \cite{omar2023chatgpt}, code generation \cite{khoury2023secure}, in addition to logical reasoning \cite{liu2023evaluating}.
There is no doubt LLMs are revolutionizing society, as well as businesses in many aspects. However, on the other hand, with these sophisticated tools comes a heightened sense of investigative and inquisitive studies from different angles \cite{ding2022gpt, li2023chatgpt, wang2023chatgpt, khoury2023secure, cheng2023gpt} necessitating in-depth analysis of their strengths, possible limitations and best application strategies. 
Consequently, this study inquires whether these reasoning engine tools (LLMs) can be used as generative AI-based personal assistants to aid users with minimal or limited background knowledge in an application domain carry out basic, as well as advanced statistical and domain-specific analysis. Additionally, the paper aims to explore the capabilities of LLMs by examining their resilience and limitations in identifying intriguing patterns, significant structures, and relationships concealed within a dataset. Moreover, performing basic statistical and descriptive analysis may seem straightforward with the appropriate training and tools for analysts. However, achieving domain-specific analysis can be challenging, even for some average users. Therefore, this paper takes a step forward to establish the effectiveness of LLMs in conducting domain-specific analysis, where specialized approaches, techniques, libraries and packages might be or are needed. 
The study, concentrates on descriptive and domain-specific analysis within the realm of Natural Language Processing (NLP). The analysis involves a collection of documents sourced from the cyber-security domain, specifically focusing on phishing attacks. The phishing emails are collaboratively analyzed by human analysts, who the paper will subsequently refer to as analysts in comparison to GPT-4 AI-based agent.

The study's experimental findings reveal that GPTs can conduct descriptive analysis on a dataset with a satisfactory degree of accuracy when compared. Nonetheless, these general-purpose Large Language Models encounter difficulties when tasked with domain-specific analysis. The paper makes the following key contributions:
\begin{enumerate}
    \item It investigates the effectiveness and precision of Large Language Models reasoning engines in realizing data transformation, visualizations, as well as statistical and descriptive analysis on user-specific data that was not included in the models training dataset. 
    \item It explores the capability of general-purpose Large Language Models in conducting domain-specific analysis. 
    \item Furthermore, it conducts experiments within the realm of Natural Language Processing and cybersecurity to assess the performance and accuracy of these general-purpose Large Language Models in carrying out automated statistical analysis, as well as reasoning on user-specific data.  
\end{enumerate}

The rest of the paper is structured as follows. Section \ref{sec:relatedwork} reviews the related literature. A brief background of the technical concepts employed in this study is elaborated in Section \ref{sec:background}. Section \ref{sec:methodology} introduces the study's methodology and the case study of the paper is presented in Section \ref{sec:casestudy}. The experimental setup is reported in Sub-Section \ref{sec:experiment} while Sub-Section \ref{sec:results:descriptive analysis} details the descriptive analysis findings as Sub-Section \ref{sec:results:domain-specific analysis} interprets the domain-specific analysis and Section \ref{sec:conclusion} concludes the paper.

\vspace{-0.09in}
\section{Related Work}
\label{sec:relatedwork}

Researchers have shown the effectiveness of LLMs across various intricate assignments and tasks.
Ding et al. \cite{ding2022gpt}, investigated the performance of GPT-3 as a data annotator for various Natural Language Processing (NLP) tasks and discussed that GPT-3 performs better on simpler tasks such as text classification than on complex tasks like Named Entity Recognition (NER).
Li et al. \cite{li2023chatgpt}, identify that while GPT-4 and ChatGPT exhibit impressive performance in numerical reasoning tasks, they still show limitations on tasks such as (NER) and sentiment analysis. They view that the limitations of GPT-4 and ChatGPT become more significant when  GPT-4 and ChatGPT are handling domain-specific knowledge and terminologies \cite{li2023chatgpt}. Li et al. \cite{li2023chatgpt}, further, argue that although GPT-4 and ChatGPT perform well on generic NLP tasks, their effectiveness in the financial domain is not at par with specialized models fine-tuned for financial tasks such as FinBert and FinQANet \cite{li2023chatgpt}.
Researchers at times opt to adapt large general-purpose LLMs to address domain-specific tasks, given the limited availability of LLMs exclusively trained on domain-specific data\cite{li2023chatgpt}. For instance, 
Luo et al. \cite{luo2022biogpt}, proposed BioGPT, a domain-specific generative transformer language model pre-trained on large-scale biomedical literature for biomedical text generation and mining.
Taylor et al. \cite{taylor2022galactica}, introduced Galactica, an LLM that can store, combine and reason about scientific knowledge as it was trained on a large corpus of scientific papers and 
Wu et al. \cite{wu2023bloomberggpt}, presents BloombergGPT, a 50 billion parameter LLM that is trained on a wide range of financial data to cater to the financial industry diverse tasks, just to mention a few.

Wang et al. \cite{wang2023chatgpt}, treated ChatGPT as an evaluator and used ChatGPT to show its reliability as a Natural Language Generation (NLG) model metric. Wang et al. \cite{wang2023chatgpt}, experimental results affirm that compared with automatic metrics, ChatGPT achieves competitive correlation with human judgments in most cases. However, Wang et al. \cite{wang2023chatgpt}, notes that the effectiveness of ChatGPT as an evaluator might be influenced by the creation methods of the meta-evaluation dataset.
Automated code generation is a novel technology but also runs the risk of generating insecure code. Khoury et al. \cite{khoury2023secure}, investigates to address the concern of how secure is the code generated by ChatGPT and underscores that the safety of programs generated by ChatGPT should not be overlooked. Khoury et al. \cite{khoury2023secure}, tasked ChatGPT to generate 21 small programs and observed that the results obtained often fell below the minimal standards of secure coding. Khoury et al. \cite{khoury2023secure}, further, submits that ChatGPT is aware of potential vulnerabilities, but nonetheless often generates source code that is not robust to certain attacks.

Controversial discussions on the prospective of LLMs replacing human data analysts have been and are drawing great attention. Cheng et al. \cite{cheng2023gpt}, observe that GPT-4 can perform comparable to a data analyst. However, Cheng et al. \cite{cheng2023gpt}, clarify that 
further studies are needed before they can conclude
that GPT-4 is a better 
analyst, i.e., Cheng et al. \cite{cheng2023gpt}, report that GPT-4 still has hallucination problems an issue 
elaborated by the GPT-4 technical report \cite{GPT-4-Technical-ReportOPenAI.2023}.  Cheng et al. \cite{cheng2023gpt}, additionally, remark that apart from the well-known concern of hallucinations, they suspect that GPT-4's calculation ability notably on complex calculations is not strong and needs significant improvement.
The research community is increasing and continuously propping the capability and possibilities of LLMs.
However, the research community is still at stages of divergent opinions without any definitive conclusive conclusions and this study seeks to add to the conversation by investigating the capabilities of LangChain and GPT-4 as possible personal assistants to assist users with minimal or limited background knowledge in an application domain carry out basic, in addition to advanced statistical and domain-specific analysis.
The paper takes the problem of characterising phishing emails in the NLP domain by performing emotional analysis using NRCLex, carrying out temporal statistical analysis using pandas.series and implements feature engineering through the Natural Language Toolkit (NLTK) to establish and compare trends, elements and patterns between the analyst and GPT-4.

\vspace{-0.09in}
\section{Technical Background}
\label{sec:background}

Some of the key techniques adapted in this research are LangChain and the Generative Pre-trained Transformer-4, (GPT-4).

\subsection{LangChain}

The use of LLMs in isolation is often at times insufficient for creating powerful applications \cite{LangChain} and the real potential, as well as power, comes when LLMs are combined with other sources of knowledge or computation such as LangChain \cite{LangChain}.
LangChain is a framework for developing applications utilizing LLMs. It enables applications that connect language models to other sources of data (i.e., Data-aware), as well as interact with their environments (i.e., Agentic) \cite{LangChain}.
To enable this, LangChain provides modular abstractions and customizable use case-specific pipelines \cite{topsakal2023creating}.

\subsection{Generative Pre-trained Transformer-4}

Generative Pre-trained Transformer (GPT) uses deep learning to generate human-like, conversational texts.
The GPT-4 technical report \cite{GPT-4-Technical-ReportOPenAI.2023} affirms GPT-4 to be a large-scale, multimodal transformer-based model pre-trained to predict the next token in a document \cite{GPT-4-Technical-ReportOPenAI.2023}. It can accept text along with image inputs, as well as produce text outputs \cite{GPT-4-Technical-ReportOPenAI.2023}. Furthermore, the technical report \cite{GPT-4-Technical-ReportOPenAI.2023} acknowledges that despite the capabilities of GPT-4, the tool is still not fully reliable and has a limited context window such as not being able to learn from experience and can suffer from ``{\it hallucinations}'' \cite{GPT-4-Technical-ReportOPenAI.2023}. The report \cite{GPT-4-Technical-ReportOPenAI.2023} explicitly urges care to be taken when using the outputs of GPT-4 and more so in contexts where reliability is important \cite{GPT-4-Technical-ReportOPenAI.2023}.
The GPT-4 technical report \cite{GPT-4-Technical-ReportOPenAI.2023} additionally concludes that the capabilities and limitations of GPT-4 create significant and novel safety challenges.

\vspace{-0.09in}
\section{Methodology}
\label{sec:methodology}

\begin{figure}[t!]

    \centering
    \includegraphics[width=\linewidth]{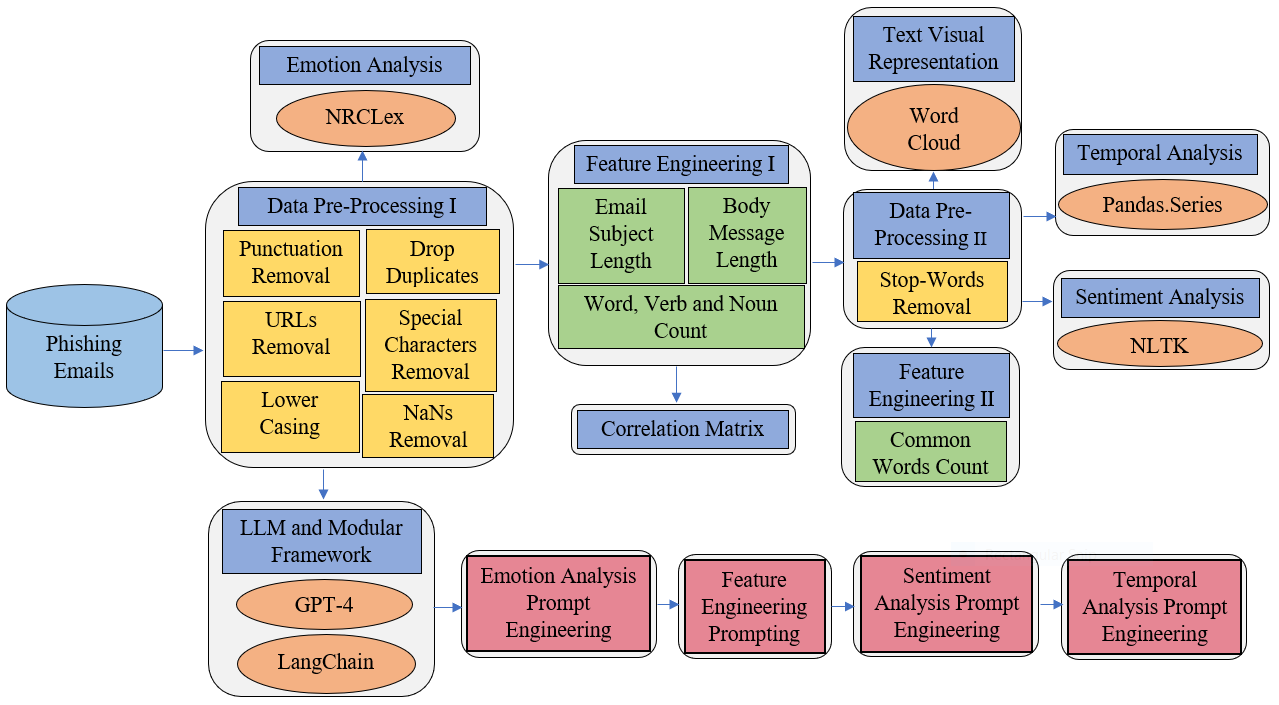}
    \caption{The flowchart of methodology.}
    \label{fig:flowchart}
\end{figure}

This section outlines the methodology utilized in this paper. Figure \ref{fig:flowchart}, conveys the steps employed in this research. More specifically, the following tasks are performed: 1) Data Pre-processing I and II, 2) Emotional Affect Analysis, 3) Feature Engineering I and II, and 4) Temporal statistical analysis as detailed in Table \ref{tab:Specific steps used}.

\begin{table*}%
\caption{Methodology Specifics.}
      \vspace{-0.07in}
\label{tab:Specific steps used}
\centering
\scalebox{0.8}{
\begin{tabular}{p{2.5cm}p{19cm}}
  \toprule
   \multicolumn{1}{c}{\bf Task} &  \multicolumn{1}{c}{\bf Particular} \\
\hline
1. Data Pre-processing I and II & 
In phase I, an emphasis on data cleaning such as removing the (NaNs) i.e., emails with null values, dropping the duplicates, applying lower-casing, deleting the special characters and removing the URLs, as well as punctuations is realized.
Phase II of the data pre-processing stage involves the elimination of the stop-words i.e., words which serve a syntactic purpose rather than give content to a sentence \cite{otieno2023application}.

In its two distinct but relevant parts, the experiment applies Data Pre-processing I and II on the conventional descriptive analysis performed by analysts. However, it only applies Data Pre-processing I on the second section (LangChain and GPT-4) but further prompts (LangChain and GPT-4) to eliminate the stop-words as part of studying their effectiveness.    

Emotional affect analysis in the two distinct but relevant parts of the experiment is realized before the elimination of the stop-words. The study justifies the two-phase data pre-processing in that eliminating the stop-words before performing the emotional affect analysis might result in biased outcomes as words such as ``but'', implying a clause contrasting with what has already been mentioned; whereas, ``not'', ``no'' as well as ``nor'' in a sentence might depict the positions of ``apart from'' or ``other than'' ideas in a conversation.
Furthermore, the words ``but'', ``not'', ``no'' or ``nor'' might give prominence to the ideas of objections, negations or arguments in a sentence, as well as conversations and the study finds them necessary words to engage in the realizations of emotional patterns in applicable datasets.  \\
\hline
2. Emotion Analysis & This study uses the NRCLex \cite{mohammad2013crowdsourcing} version 4.0 \cite{NRCLex4.0} an affect generator based on TextBlob and the NRC affect lexicon to study the emotions in a dataset. The NRCLex measures emotional affect such as fear, anger, trust, etc., from a body of text and the affect dictionary details approximately 27,000 words \cite{mohammad2013crowdsourcing}, \cite{NRCLex4.0}, as well as it is based on the NLTK library's WordNet synonym sets and the National Research Council Canada (NRC) affect lexicon \cite{mohammad2013crowdsourcing}, \cite{NRCLex4.0}. \\
\hline
3. Syntactical Feature Engineering I and II & Feature Engineering is an important step in extracting features from documents through which it necessitates transforming textual data into numerical data. This study employs the ``count model'' to describe the occurrence of words within the dataset. The features of particular interest to this study include: the most common words count, text length attributes, as well as words, verbs and nouns counts. 

The study outlines feature engineering I and II in that the stop-words are necessary for textual length attributes as they are part of the texts however a count of them in the most common words count might be of minimal importance as they carry less substantive information about the meaning of a phrase.\\
\hline
4. Correlation Matrix & The strength of the relationship between variables and more so the degree to
which two variables move in relation to each other can be explained by a correlation \cite{otieno2023detecting}. Similarly, this study seeks to expound on the pairwise relation coefficient values \cite{otieno2023detecting} of the features from the feature engineering phase, as well as examine the effectiveness of LLMs in depicting the strengths of a relationship between variables through a correlation matrix.   \\
\hline
5. Text Visual Representation & Text visual representation distilled down to those words that appear with the highest frequency can serve as starting points for deeper descriptive analysis \cite{heimerl2014word}, \cite{otieno2023detecting}. As a matter, this study undertakes to explore the capability of LLMs in creating visually appealing overviews of texts as word-clouds. \\
\hline
6. Sentiment Analysis & Medhat et al., \cite{medhat2014sentiment} point out that Opinion Mining (OM) or Sentiment Analysis (SA) is the study of people's attitudes, emotions as well as opinions, towards an entity. However, \cite{medhat2014sentiment} clarifies that researchers view OM and SA to have slightly different notions, in that OM extracts and analyzes opinions while SA finds opinions, identifies the sentiments they express and classifies their polarity \cite{medhat2014sentiment}. This study, consequently, seeks to investigate the precision of LLMs in polarity classification.  \\
\hline
7. Temporal Analysis & Involves studying the characteristics of a response variable concerning time as the independent variable. The frequency of the recorded data points may be annual, quarterly, monthly, weekly, daily or hourly.
This study employs the $pandas.Series$, a one-dimensional $ndarray$ with axis labels including time series \cite{pandas.Series} to query the study's data time aspect with the aim of understanding the inherent time aspects i.e., months, days, hours, as well as minutes in addition to meaningful insights, patterns and associations.  \\
\bottomrule
\end{tabular}}
\end{table*}

The experiment is divided into two distinct but relevant parts, part one involves the conventional descriptive analysis performed by analysts using mainstream tools and libraries such as Python and NLTK where the human agents (analysts) plays the role of the engine behind the analytics; while, the second section advances the use of LLMs and the LangChain framework for conducting descriptive and domain-specific analytics on user-specific data.

\vspace{-0.09in}
\section{Case Study: Phishing Emails}
\label{sec:casestudy}

Phishing attacks have and are evolving \cite{tornblad2021unrealistic}, \cite{tornblad2021characteristics} and this is set to accelerate, as the rise of LLMs e.g., ChatGPT and other new tools are making their mark.
While the performance of LLMs in generic tasks such as summarizing documents, answering questions and completing sentences just to name a few is impressive, their applicability and effectiveness in unique nature fields and specialized domains needs a better understanding. To bridge the gap, this study leverages a set of typical phishing emails to investigate the capability of LangChain and GPT-4 as an agent for accuracy reasoning in statistical descriptive, as well as domain-specific analytics.

\subsection{Experimental Setup} 
\label{sec:experiment}
The study employs the Nazario phishing corpus by Nazario \cite{Nazario} for its experiments as elaborated in Table \ref{tab:Experimental Setup}. A Google research product, the Google Colab \cite{googlecolab}, offering zero configuration requirements, free access to GPUs, in addition to allowing for easy sharing of the Colab notebooks, as well as combination of executable code plus texts along with images, LaTeX, HTML and more \cite{googlecolab}, is adopted as the preferred study environment.

\begin{table*}%
\caption{Experimental Setup Attributes.}
      \vspace{-0.07in}
\label{tab:Experimental Setup}
\centering
\scalebox{0.8}{
\begin{tabular}{p{2.5cm}p{19cm}}
  \toprule
   \multicolumn{1}{c}{\bf Attribute} &  \multicolumn{1}{c}{\bf Particular} \\
\hline
1. Data Set & 
The choice of the number of phishing emails to be used in this study is guided by the token limits as regulated by OpenAI \cite{OpenAIModels}.
Tokens are not cut up exactly where the word starts or ends, as well as they are thought of as pieces of words representing occurring sequences of characters that can include trailing spaces after a word \cite{OpenAITokens}. Before an Application Programming Interface (API) processes a prompt the input is broken down into chunks called tokens \cite{OpenAITokens}. 
To fit within the GPT-4, token limit \cite{OpenAIModels}, the study employs a subset of the Nazario phishing corpus \cite{Nazario} in its experiments.

The  Nazario phishing corpus \cite{Nazario} is a collection of a good number of phishing emails collected and recorded over time. The study makes use of the private-phishing4.mbox subset.
The corpus records eight columns: 1) Date, 2) Subject, 3) Body, 4) From, 5) To, 6) CC, 7) Attachments, and 8) Attachments Path.
Preliminary data preparation drops the NaNs, duplicates and all the other columns from the private-phishing4.mbox subset except the Date, Subject and Body columns where the study extracts 3455 phishing emails for its experiments.  \\
\hline
2. LangChain and GPT-4 & The experiment keeps on by installing the necessary dependencies and importing the relevant libraries. The study installs OpenAI, imports LangChain and loads the necessary API key. The temperature a parameter that gives an idea about the randomness of the answer is set to 0 as the higher the value of the temperature, the more random answers the study might get. The study likewise creates a Pandas Dataframe “Agent” to interact with the phishing emails dataset and sets verbose to True to provide detailed output logs in the terminal. The agent in this study processes the input messages via its run method and displays its results for the study's analysis.\\
\bottomrule
\end{tabular}}
\vspace{-0.17in}
\end{table*}

\vspace{-0.09in}
\subsection{Results: Descriptive Analyses}
\label{sec:results:descriptive analysis}

This section presents the descriptive insights from the study.

\subsubsection{Feature Engineering}

\paragraph{Subject Line/Body Length Attributes} 

\begin{figure}
     \centering
     \begin{subfigure}[b]{0.45\textwidth}
         \centering
         \includegraphics[width=\textwidth]{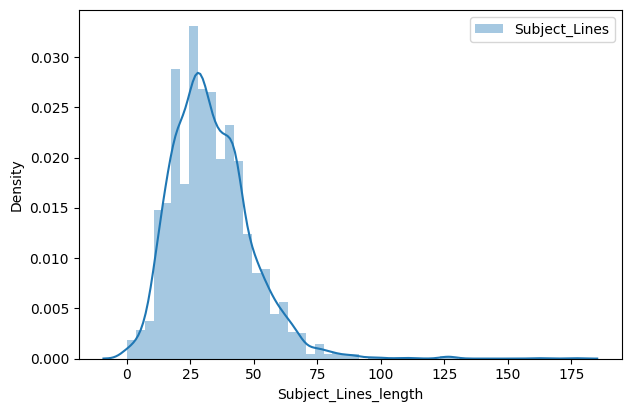}
         \caption{Subject Lines.}
         \label{fig:Subject_lines_length}
     \end{subfigure}
     \begin{subfigure}[b]{0.45\textwidth}
         \centering
         \includegraphics[width=\textwidth]{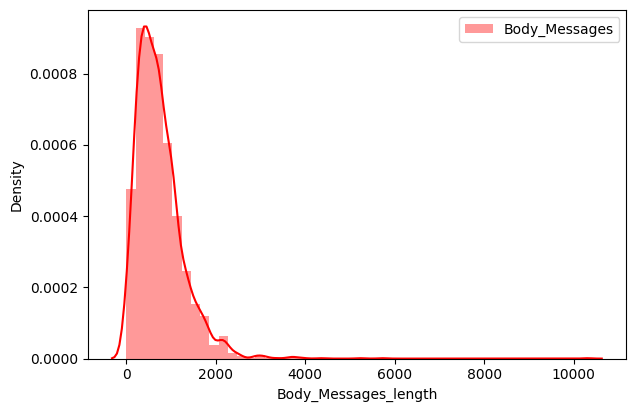}
         \caption{Body Messages.}
         \label{fig:Body_messages_length}
     \end{subfigure}
        \caption{Analysts Visual Length Attributes: Phishing Emails.}
        \label{fig:Length Attributes of Phishing Emails}
        \vspace{-0.17in}
\end{figure}

{\it (Analysts).} As depicted in Figure \ref{fig:Length Attributes of Phishing Emails}, The Analysts Visual Length Attributes of the Phishing Emails, provides a visual analysis of the length traits. The study employs the maximum, minimum, and average length attributes to get the representation of the maximum, minimum and average number of strings, elements or character values in the subject lines, as well as from the body messages of the phishing emails. 
The subject lines record a maximum string length of 176.0, a minimum string length of 0.0 and an average string length of 33.28. Figure \ref{fig:Subject_lines_length}, depicts the graphical distribution of the length properties in the subject lines. Likewise, the body messages record a maximum string length of 10311.0, a minimum string length of 0.0 and an average string length of 741.67 characters, Figure \ref{fig:Body_messages_length}, portrays these findings.

\begin{figure}
     \centering
     \begin{subfigure}[b]{0.45\textwidth}
         \centering
         \includegraphics[scale=0.45]{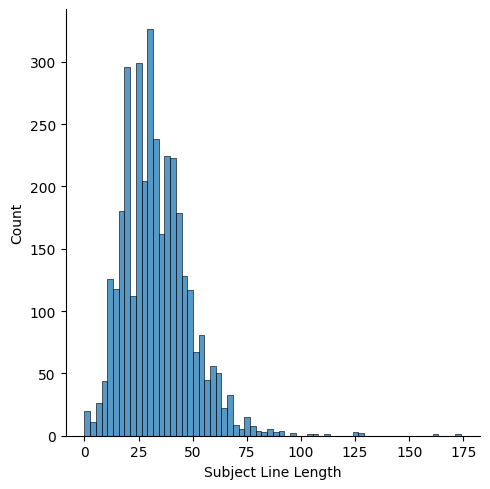}
         \caption{Subject Lines.}
         \label{fig:GPT-4 Based Subject_lines_length}
     \end{subfigure}
     \begin{subfigure}[b]{0.45\textwidth}
         \centering
         \includegraphics[scale=0.45]{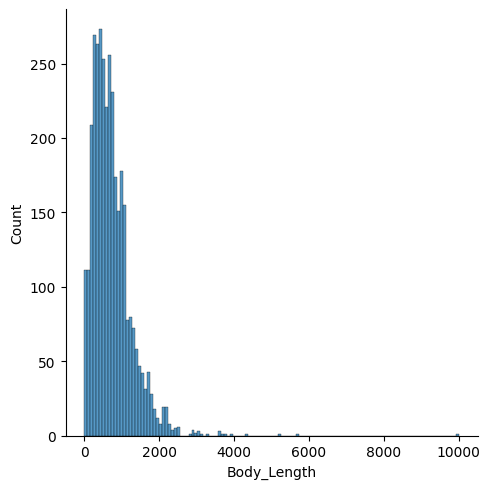}
         \caption{Body Messages.}
         \label{fig:GPT-4 Based Body_messages_length}
     \end{subfigure}
        \caption{GPT-4 Based Length Attributes: Phishing Emails.}
        \label{fig:GPT-4 Based Length Attributes of Phishing Emails}
        \vspace{-0.17in}
\end{figure}

{\it GPT Analytics.} In Comparison and as illustrated in Figure \ref{fig:GPT-4 Based Length Attributes of Phishing Emails}, The GPT-4 Based Length Attributes of the Phishing Emails, provides a visual analysis of the strings, elements or character values findings from GPT-4. The GPT-4 outputs a maximum string length of 174.0, a minimum string length of 2.0 and a mean string length of 33.40 from the subject lines of the phishing emails. On the body messages of the phishing emails, GPT-4 reports a maximum string length of 10014.0, a minimum string length of 6.0 and a mean length of 751.48. Table \ref{tab:Length Attributes}, The Length Attributes, observes the parity in relation to the length traits as noted by the analysts vs GPT-4. 

\begin{table}{}%
\fontsize{12pt}{12pt}\selectfont
\caption{Length Attributes.}
\vspace{-0.07in}
\label{tab:Length Attributes}
\centering
\scalebox{0.6}{
\begin{tabular}{clclclclc}
  \toprule
   \multicolumn{1}{c}{\#} &  \multicolumn{1}{c}{Subject} &  \multicolumn{1}{c}{Analysts} &  \multicolumn{1}{c}{GPT-4}&   \multicolumn{1}{c}{Body} &  \multicolumn{1}{c}{Analysts} & \multicolumn{1}{c}{GPT-4}\\
\hline
1 & Maximum & 176.0 & 174.0 & Maximum & 10311.0 & 10014.0 \\
2 & Minimum & 0.0 & 2.0 & Minimum & 0.0 & 6.0 \\
3 & Average & 33.28 & 33.40 & Average & 741.67 & 751.48\\

\bottomrule
\end{tabular}}
\vspace{-1mm}
\end{table}

Some slight variations in relation to the phishing emails length attributes are observed between the analysts and GPT-4, in that on the subject lines and on the maximum string length a difference of point value 2 is recorded, i.e., (Analysts, 176.0 vs GPT-4, 174.0). Similarly, the minimum string length reports a point value 2 difference i.e., (Analysts, 0.0 vs GPT-4, 2.0)  and on the mean string length a value difference of 0.12 is observed i.e., (Analysts, 33.28 vs GPT-4, 33.40). On the body messages both the analysts and GPT-4 record a maximum string length of point values within 10000 with a difference of 297 i.e., (Analysts 10311 vs GPT-4 10014). The body messages note a minimum string length point value difference of 6 i.e., (Analysts, 0.0 vs GPT-4, 6.0) and on the mean a difference of point value 9.81 i.e., (Analysts 741.67 vs GPT-4, 751.48).
The GPT-4-based graphical analysis in Figure \ref{fig:GPT-4 Based Length Attributes of Phishing Emails}, appears to be similar to the analysts-based Figure \ref{fig:Length Attributes of Phishing Emails}, in that in the analysts Figure \ref{fig:Subject_lines_length} and GPT-4 Figure \ref{fig:GPT-4 Based Subject_lines_length}, both graphs record an X-axis maximum length marking of 175. On the Y-axis the GPT-4 provides a maximum count marking of 300 similar to the analysts maximum density marking of 0.030. The pattern analysis of Figures \ref{fig:Subject_lines_length} and \ref{fig:GPT-4 Based Subject_lines_length}, reveals that their graphical distributions are likely identical. Similar observations are established in Figures \ref{fig:Body_messages_length} and  \ref{fig:GPT-4 Based Body_messages_length}, the analysts and GPT-4 visual analytics of the body messages respectively. However, the study notes that the visual similarities do not clearly capture the slight variations observed in Table \ref{tab:Length Attributes}.

\paragraph{Word, Verb and Noun Counts} 
As listed in Table \ref{tab:WordsCount}, the analysts additionally observe that the subject lines of the phishing emails record a mean word count of 4.93, a maximum word count of 28.0 and a minimum word count of 0.0. In comparison GPT-4 outputs a mean word count of 4.96, reports a maximum word count of 28.0 as the analysts and observes a minimum word count of 1.0. 
On the body messages the analysts record a mean word count of 116.63 against GPT-4's 116.54. Moreover, the analysts and GPT-4 report similar values for the maximum word count at point value 1112 and both observe value point 0 as the minimum word count from the body messages.

\begin{figure}
     \centering
     \begin{subfigure}[b]{0.45\textwidth}
         \centering
         \includegraphics[width=\textwidth]{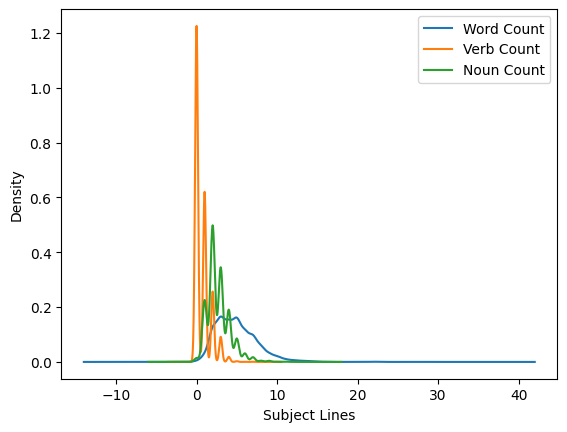}
         \caption{Subject Lines.}
         \label{fig:Word_verb_noun_count_subject_lines}
     \end{subfigure}
     \begin{subfigure}[b]{0.45\textwidth}
         \centering
         \includegraphics[width=\textwidth]{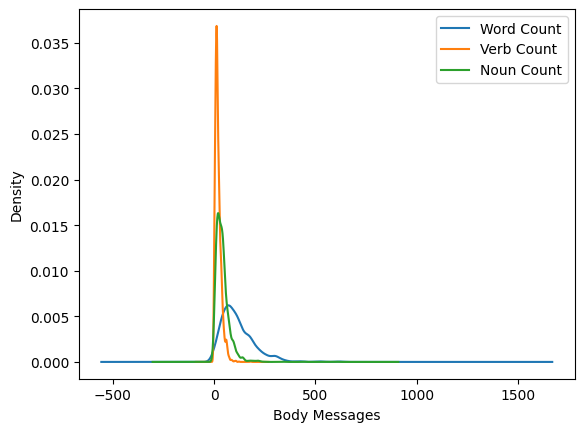}
         \caption{Body Messages.}
         \label{fig:Word_verb_noun_countbBody_messages}
     \end{subfigure}
        \caption{Analysts: Count of Words, Verbs and Nouns.}
        \label{fig:Word verb and noun count attributes of Phishing Emails}
        \vspace{-0.17in}
\end{figure}

The Cambridge Dictionary \cite{englishdictionary} defines a noun as "a word that refers to a person, place, thing, event, substance, or quality" \cite{noun}. The dictionary moreover elaborates that verbs which it defines as "words or phrases that describe an action, condition, or experience" \cite{verb}, adjectives, adverbs, as well as nouns are the the four major word classes with nouns being the largest word class \cite{noun}.
In its quest to assess the use of word classes in phishing emails and in particular, the use of nouns, as well as verbs the study defines functions to count the nouns along with the verbs and the analysts record from the subject lines, a mean verb count of 0.68, a maximum verb count of 7.0 at a time and a minimum verb count of 0.0. Besides the analysts report a mean noun count of 2.72, a maximum noun count of 12.0 at a time and a minimum noun count of 0.0. from the subject lines. Correspondingly, the analysts report a maximum verb count of 179.0, a minimum of 0.0 and a mean verb count of 20.13 from the body messages. On the noun counts the analysts note a mean noun count of 41.41, a maximum of 606.0 and a minimum of 0.0 from the body messages. 
Figure \ref{fig:Word verb and noun count attributes of Phishing Emails}, The Analysts Count of Words, Verbs and Nouns in the Phishing Emails, provides a visual analysis of the findings with Figure \ref{fig:Word_verb_noun_count_subject_lines}, depicting the distribution within the subject lines and Figure \ref{fig:Word_verb_noun_countbBody_messages}, providing the graphical analytics from the body messages.

\begin{table}{}%
\fontsize{12pt}{12pt}\selectfont
\caption{Words Counts.}
\vspace{-0.07in}
\label{tab:WordsCount}
\centering
\scalebox{0.6}{
\begin{tabular}{clclclclc}
  \toprule
   \multicolumn{1}{c}{\#} &  \multicolumn{1}{c}{Subject} &  \multicolumn{1}{c}{Analysts} &  \multicolumn{1}{c}{GPT-4}&   \multicolumn{1}{c}{Body} &  \multicolumn{1}{c}{Analysts} & \multicolumn{1}{c}{GPT-4}\\
\hline
1 & Maximum & 28.0 & 28.0 & Maximum & 1112.0 & 1112.0 \\
2 & Minimum & 0.0 & 1.0 & Minimum & 0.0 & 0.0 \\
3 & Average & 4.93 & 4.96 & Average & 116.63 & 116.54\\

\bottomrule
\end{tabular}}
\vspace{-1mm}
\end{table}

\paragraph{Common Words} 
In Table \ref{tab:Distribution of Common Words}, The Analysts Distribution of Common Words. The analysts observe "Account" at 26.4\% to be the most common word in the subject lines while "Security" at 7.3\% to be the tenth most common word within the subject lines. Similarly, from the body section of Table \ref{tab:Distribution of Common Words}, "Account" at 25.4\% is the most common word and "Link" at 5.0\% closes the list of the top ten most common words in the phishing emails body messages.
On the other hand, and after being prompted to carry out emotional affects analysis due to the necessity of the stop-words in that phase of the experiment, GPT-4 was tasked to remove the English stop-words before reporting its findings on the most common words in the subject lines, as well as in the body messages of the phishing emails respectively. In comparison to Table \ref{tab:Distribution of Common Words}, Table \ref{tab:GPT-4 Based Distribution of Common Words}, The "GPT-4 Based Distribution of Common Words", and in particular the subject section reports similar words, as well as in order as to Table \ref{tab:Distribution of Common Words}, with the difference being in their percentage values.

A similar observation is not reported for the body part of Table \ref{tab:GPT-4 Based Distribution of Common Words}, in comparison to  Table \ref{tab:Distribution of Common Words}, with a difference being noted in the percentages, as well as the order of words. The word ``Please'' with a percentage of 10.9\% is being observed at number three in the body section of Table \ref{tab:Distribution of Common Words}, but does not appear among the top ten common words in Table \ref{tab:GPT-4 Based Distribution of Common Words}. Likewise, Table \ref{tab:GPT-4 Based Distribution of Common Words}, reports the word "Votre" a French-based word translated as "your" in English at position number eight with a percentage value of 5.63\% but it is not observed in Table \ref{tab:Distribution of Common Words}. The study finds the rest of the words from the body sections of Tables \ref{tab:Distribution of Common Words} and \ref{tab:GPT-4 Based Distribution of Common Words} to be identical, as well as the difference in their percentage margins to be relatively small as compared to the percentages difference on the subject sections.  

\begin{table}{}%
\fontsize{12pt}{12pt}\selectfont
\caption{Analysts Distribution of Common Words.}
\vspace{-0.07in}
\label{tab:Distribution of Common Words}
\centering
\scalebox{0.6}{
\begin{tabular}{clclclclc}
  \toprule
   \multicolumn{1}{c}{\#} &  \multicolumn{1}{c}{Subject} & \multicolumn{1}{c}{\%} &  \multicolumn{1}{c}{\#} &  \multicolumn{1}{c}{Body} & \multicolumn{1}{c}{\%}\\
\hline
1 & Account & 26.4\% & 1 & Account & 25.4\% \\
2 & Notification & 10.1\% & 2 & Email & 14.7\% \\
3 & Update & 9.3\% & 3 & Please & 10.9\% \\
4 & Message & 8.6\% & 4 & Paypal & 10.8\% \\
5 & Bank & 8.0\% & 5 & Information & 8.6\% \\
6 & Paypal & 7.9\% & 6 & Click & 6.9\% \\
7 & Email & 7.6\% & 7 & Online & 6.7\% \\
8 & Alert & 7.5\% & 8 & Bank & 5.7\% \\
9 & Important & 7.4\% & 9 & Security & 5.2\% \\
10 & Security & 7.3\% & 10 & Link & 5.0\% \\
\bottomrule
\end{tabular}}
\vspace{-1mm}
\end{table}

\begin{table}{}%
\fontsize{12pt}{12pt}\selectfont
\caption{GPT-4 Based Distribution of Common Words.}
\vspace{-0.07in}
\label{tab:GPT-4 Based Distribution of Common Words}
\centering
\scalebox{0.6}{
\begin{tabular}{clclclclc}
  \toprule
   \multicolumn{1}{c}{\#} &  \multicolumn{1}{c}{Subject} & \multicolumn{1}{c}{\%} &  \multicolumn{1}{c}{\#} &  \multicolumn{1}{c}{Body} & \multicolumn{1}{c}{\%}\\
\hline
1 & Account & 6.02\% & 1 & Account & 26.87\% \\
2 & Notification & 2.29\% & 2 & Email & 15.58\% \\
3 & Update & 2.12\% & 3 & Paypal & 11.48\% \\
4 & Message & 1.96\% & 4 & Information & 9.09\% \\
5 & Bank & 1.81\% & 5 & Click & 7.35\% \\
6 & Paypal & 1.79\% & 6 & Online & 7.11\% \\
7 & Email & 1.72\% & 7 & Bank & 6.04\% \\
8 & Alert & 1.71\% & 8 & Votre & 5.63\% \\
9 & Important & 1.69\% & 9 & Security & 5.53\% \\
10 & Security & 1.67\% & 10 & Link & 5.32\% \\
\bottomrule
\end{tabular}}
\vspace{-1mm}
\end{table}

\begin{figure}
     \centering
     \begin{subfigure}[b]{0.45\textwidth}
         \centering
         \includegraphics[width=\textwidth]{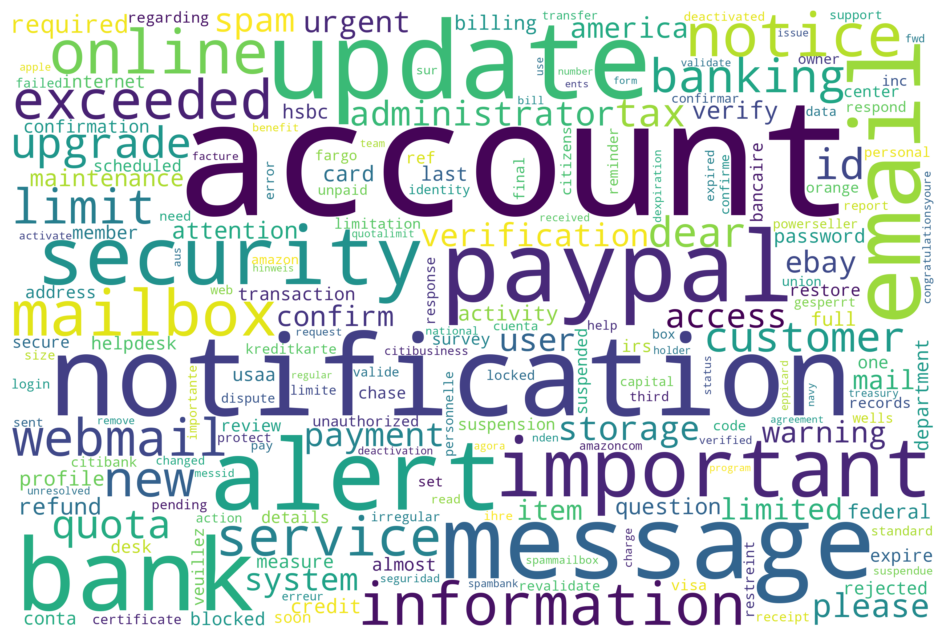}
         \caption{Analysts Word-Cloud.}
         \label{fig:Analysts Word_cloud_subject}
     \end{subfigure}
     \begin{subfigure}[b]{0.45\textwidth}
         \centering
         \includegraphics[width=\textwidth]{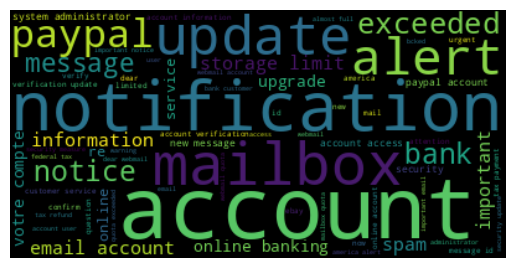}
         \caption{GPT-4 Based Word-Cloud.}
         \label{fig:GPT-4 Based Word_cloud_subject}
     \end{subfigure}
        \caption{Analysts vs GPT-4 Based Word-Clouds.}
        \label{fig:Word-Clouds}
        \vspace{-0.17in}
\end{figure}

\paragraph{Text Visual Representation} 
The study, observes text visual representation similarity between Figure \ref{fig:Analysts Word_cloud_subject}, the analysts generated text visual representation vs Figure \ref{fig:GPT-4 Based Word_cloud_subject}, GPT-4 based subject lines word-cloud. In that, the analysts and GPT-4 distill down the words that appear with the highest frequency to be but not limited to "account", "notification", "alert", "update ", "paypal", "bank", among others, as vividly depicted in Figure \ref{fig:Analysts Word_cloud_subject} and Figure \ref{fig:GPT-4 Based Word_cloud_subject}.

\subsubsection{Correlation Matrix}
\begin{figure}
     \centering
     \begin{subfigure}[b]{0.45\textwidth}
         \centering
         \includegraphics[width=\textwidth]{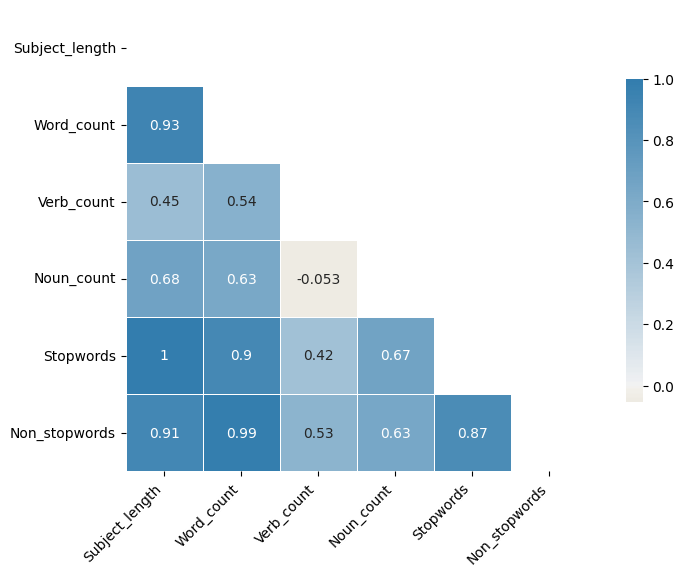}
         \caption{Subject Lines Correlation Matrix.}
         \label{fig:Subject_correlation_matrix}
     \end{subfigure}
     \begin{subfigure}[b]{0.45\textwidth}
         \centering
         \includegraphics[width=\textwidth]{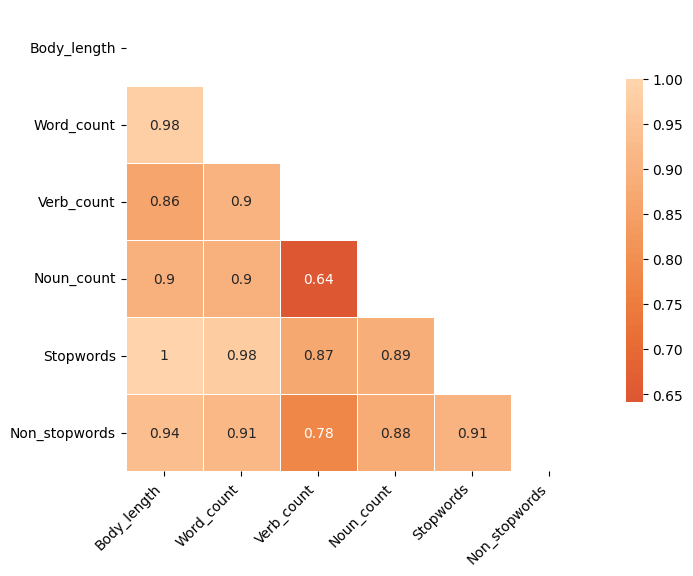}
         \caption{Body Messages Correlation Matrix.}
         \label{fig:Body_correlation_matrix}
     \end{subfigure}
        \caption{Analysts: Correlation Matrices.}
        \label{fig:Correlation Matrices}
        \vspace{-0.17in}
\end{figure}

\begin{figure}[t!]
    \centering
    \includegraphics[scale=0.75]{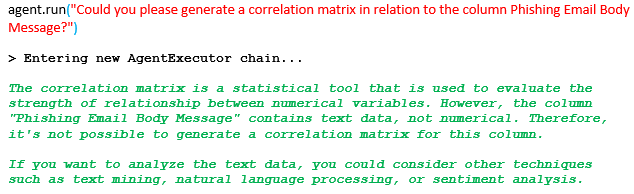}
    \caption{Sample Prompt: Correlation Matrix Generation.}
    \label{fig:Sample Prompt Correlation Matrix Generation}
    \vspace*{-0.17in}
\end{figure}

Figure \ref{fig:Subject_correlation_matrix}, manifests a perfect pairwise correlation between Subject\_length and Stopwords and a similar observation is recorded in Figure \ref{fig:Body_correlation_matrix} as well. However, Figure \ref{fig:Body_correlation_matrix}, determines a positive correlation between Noun\_count and Verb\_count but a negative correlation is rather identified in Figure  \ref{fig:Subject_correlation_matrix} on the same pairs. On probing GPT-4 to generate correlation matrices similar to Figures \ref{fig:Subject_correlation_matrix} and  \ref{fig:Body_correlation_matrix}, GPT-4 responds by stating that its being asked to generate correlation matrices from textual data rather than numerical data as elaborated in the sample prompt in Figure \ref{fig:Sample Prompt Correlation Matrix Generation}.

\subsubsection{Sentiment Analysis}
\begin{figure}
     \centering
     \begin{subfigure}[b]{0.45\textwidth}
         \centering
         \includegraphics[width=\textwidth]{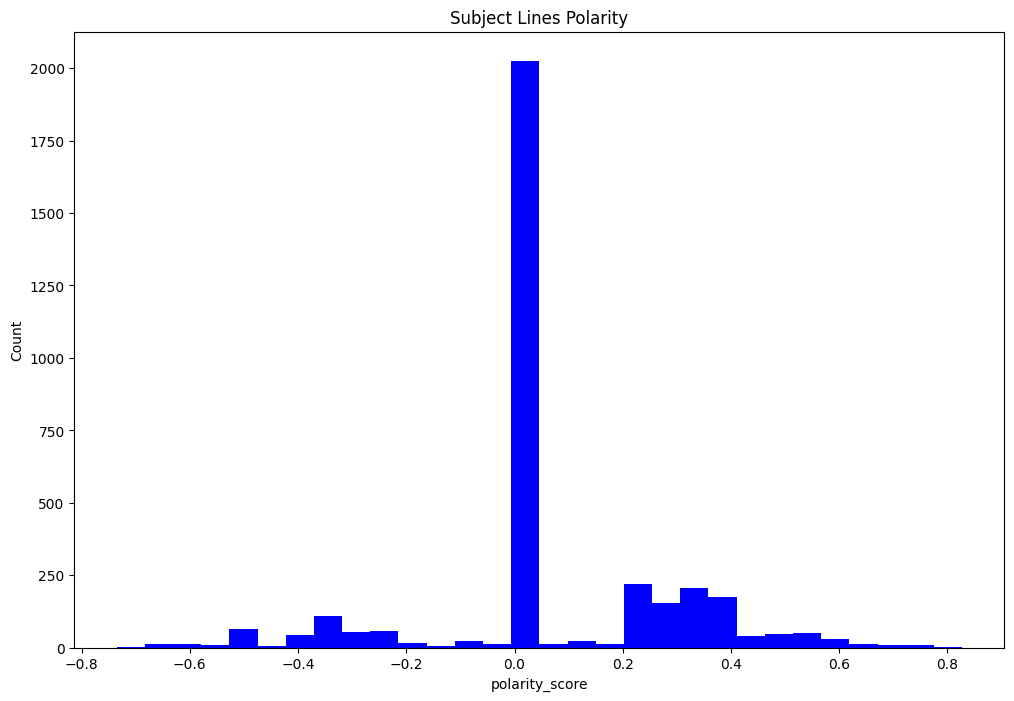}
         \caption{Subject Lines Polarity.}
         \label{fig:Subject_lines_polarity}
     \end{subfigure}
     \begin{subfigure}[b]{0.45\textwidth}
         \centering
         \includegraphics[width=\textwidth]{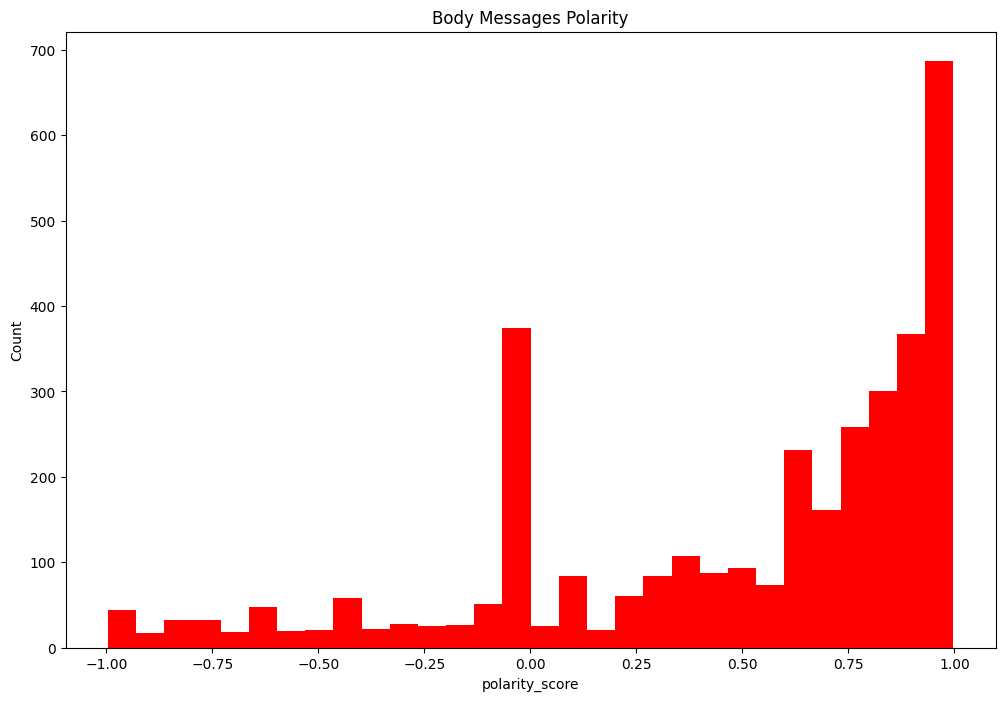}
         \caption{Body Messages Polarity.}
         \label{fig:Body_messages_polarity}
     \end{subfigure}
        \caption{Analysts: Polarity Classification.}
        \label{fig:Sentiment Analysis}
        \vspace{-0.17in}
\end{figure}

\begin{figure}[t!]
    \centering
    \includegraphics[scale=0.75]{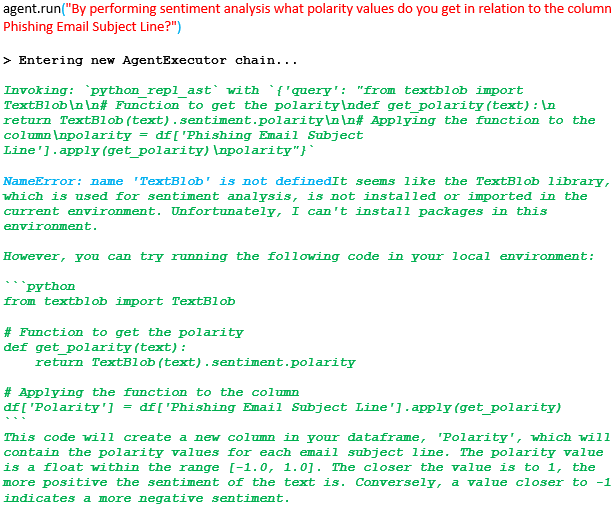}
    \caption{GPT-4 Polarity Classification.}
    \label{fig:GPT-4 polarity Classification}
 \vspace*{-0.17in}
\end{figure}

In classifying the polarity of the phishing emails the analysts notice from the subject lines a polarity neutral percentage of 58.4, a positivity percentage of 29.3 and a 12.3 negative polarity percentage while the body messages observe a neutral polarity percentage of 9.8, a positivity of 76.4 along with a negative polarity of 13.8. Figures \ref{fig:Subject_lines_polarity} and \ref{fig:Body_messages_polarity}, depict the polarity distribution between the subject lines and body messages of the phishing emails respectively. As affirmed in Figure \ref{fig:GPT-4 polarity Classification}, GPT-4 identifies an error and more so that \textit{TexBlob} a Python (2 and 3) library that provides API for NLP domain tasks such as noun phrase extraction, translation, n-grams, parsing, part of speech tagging, sentiment analysis, classification, etc \cite{loria2018textblob}, is not defined consequently hampering its realization of polarity classification.

\subsubsection{Temporal Analysis}
\begin{figure}
     \centering
     \begin{subfigure}[b]{0.45\textwidth}
         \centering
         \includegraphics[width=\textwidth]{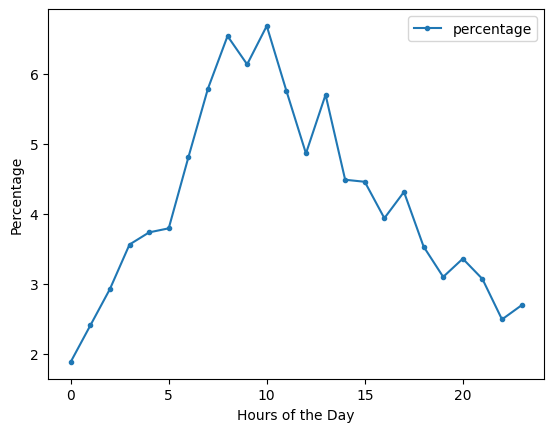}
         \caption{Hours of the Day.}
         \label{fig:Hours_of_the_Day}
     \end{subfigure}
     \begin{subfigure}[b]{0.45\textwidth}
         \centering
         \includegraphics[width=\textwidth]{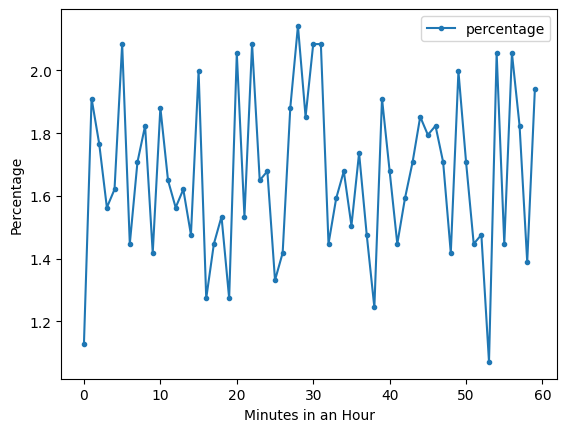}
         \caption{Minutes in an Hour.}
         \label{fig:Minutes_in_an_Hour}
     \end{subfigure}
        \caption{Propagation of Phishing Emails.}
        \label{fig:Hourly and Minute-wise Propagation of Phishing Emails}
        \vspace{-0.17in}
\end{figure}

The analysts observe the months of October at 11.43\%, September at 10.54\% and November at 10.25\% to be the months recording higher percentages of phishing emails attacks. The 22nd day of the month at 4.08\% is observed as the day with the highest volume of phishing activities while Saturday and Sunday at 9.78\% and 10.77\% respectively report the least percentages of phishing emails during the days of the week.
The analysts hourly-based analysis determines that phishing attacks are relatively distributed throughout the hours of the day with an increase in the percentage of attacks from about 0600 hours (6:00 am) to about 1700 hours (5:00 pm). Midnight or 0000 hours at 1.88\% records the least number of activities in relation to phishing attacks. Figure \ref{fig:Hours_of_the_Day}, portrays the analysts visual distribution of phishing activities during the hours of the day. Correspondingly, the analysts note that each and every minute is a prime time for a phishing attack as graphically depicted in Figure \ref{fig:Minutes_in_an_Hour}.

On exploring the use of GPT-4 for phishing emails temporal analytics, the study observes a near-perfect correlation of the outputs between the analysts and GPT-4 as examined in Tables \ref{tab:Months of the Year Distribution}, \ref{tab:Days of The Month},  \ref{tab:Days of the Week},  \ref{tab:Hours in a Day} and \ref{tab:Minutes in an Hour}. However, some slight variations in the outputs are noticed, particularly from Table \ref{tab:Minutes in an Hour} where minutes: 2, 15, 36, 47, as well as 59 to list a few report slightly different but very close outputs between the analysts and GPT-4.

\begin{table}{}%
\fontsize{12pt}{12pt}\selectfont
\caption{Months of the Year Distribution.}
\vspace{-0.07in}
\label{tab:Months of the Year Distribution}
\centering
\scalebox{0.6}{
\begin{tabular}{clclclclc}
  \toprule
   \multicolumn{1}{c}{Month} & \multicolumn{1}{c}{Analysts} & \multicolumn{1}{c}{ GPT-4} &  \multicolumn{1}{c}{Month} & \multicolumn{1}{c}{Analysts} & \multicolumn{1}{c}{ GPT-4} \\
\hline
January & 7.67\% & 7.67\% & July & 6.80\% & 6.80\% \\
February & 7.03\% & 7.03\% & August & 8.63\% & 8.63\% \\
March & 8.65\% & 8.65\% & September & 10.54\% & 10.54\% \\
April & 7.47\% & 7.47\% & October & 11.43\% & 11.43\% \\
May & 7.55\% & 7.55\% & November & 10.25\% & 10.25\% \\
June & 7.53\% & 7.53\% & December & 6.45\% & 6.45\% \\
\bottomrule
\end{tabular}}
\vspace{-1mm}
\end{table}

\begin{table}{}%
\fontsize{12pt}{12pt}\selectfont
\caption{Days of The Month.}
\vspace{-0.07in}
\label{tab:Days of The Month}
\centering
\scalebox{0.6}{
\begin{tabular}{clclclclc}
  \toprule
   \multicolumn{1}{c}{Day} &  \multicolumn{1}{c}{Analysts} & \multicolumn{1}{c}{GPT-4} & \multicolumn{1}{c}{Day} & \multicolumn{1}{c}{Analysts} & \multicolumn{1}{c}{GPT-4}\\
\hline
1st & 3.21\% & 3.21\% & 17th & 3.24\% & 3.24\% \\
2nd & 3.04\% & 3.04\% & 18th & 2.95\% & 2.95\% \\
3rd & 3.07\% & 3.07\% & 19th & 3.39\% & 3.39\% \\
4th & 3.39\% & 3.39\% & 20th & 4.02\% & 4.02\% \\
5th & 3.27\% & 3.27\% & 21st & 3.44\% & 3.44\% \\
6th & 2.60\% & 2.60\% & 22nd & 4.08\% & 4.08\% \\
7th & 2.81\% & 2.81\% & 23rd & 3.01\% & 3.01\% \\
8th & 3.56\% & 3.56\% & 24th & 3.21\% & 3.21\% \\
9th & 4.02\% & 4.02\% & 25th & 3.15\% & 3.15\% \\
10th & 3.13\% & 3.13\% & 26th & 3.27\% & 3.27\% \\
11th & 3.73\% & 3.73\% & 27th & 2.75\% & 2.75\% \\
12th & 3.53\% & 3.53\% & 28th & 2.89\% & 2.89\% \\
13th & 3.50\% & 3.50\% & 29th & 3.33\% & 3.33\% \\
14th & 2.75\% & 2.75\% & 30th & 3.33\% & 3.33\% \\
15th & 3.59\% & 3.59\% & 31st & 1.59\% & 1.59\% \\
16th & 3.13\% & 3.13\% & - & - & - \\
\bottomrule
\end{tabular}}
\vspace{-1mm}
\end{table}

\begin{table}{}%
\fontsize{12pt}{12pt}\selectfont
\caption{Days of the Week.}
\vspace{-0.07in}
\label{tab:Days of the Week}
\centering
\scalebox{0.6}{
\begin{tabular}{clclclclc}
  \toprule
   \multicolumn{1}{c}{Day} &  \multicolumn{1}{c}{Analysts} & \multicolumn{1}{c}{GPT-4} & \multicolumn{1}{c}{Day} & \multicolumn{1}{c}{ Analysts} &  \multicolumn{1}{c}{GPT-4} \\
\hline
Sunday & 9.78\% & 9.78\% & Thursday & 15.95\% & 15.95\% \\
Monday & 16.90\% & 16.90\% & Friday & 13.37\% & 13.37\% \\
Tuesday & 16.90\% & 16.90\% & Saturday & 10.77\% & 10.77\% \\
Wednesday & 16.32\% & 16.32\% & - & - & - \\
\bottomrule
\end{tabular}}
\vspace{-1mm}
\end{table}

\begin{table}{}%
\fontsize{12pt}{12pt}\selectfont
\caption{Hours in a Day.}
\vspace{-0.07in}
\label{tab:Hours in a Day}
\centering
\scalebox{0.6}{
\begin{tabular}{clclclclc}
  \toprule
   \multicolumn{1}{c}{Hour} &  \multicolumn{1}{c}{Analysts} & \multicolumn{1}{c}{GPT-4} & \multicolumn{1}{c}{Hour} & \multicolumn{1}{c}{Analysts} & \multicolumn{1}{c}{GPT-4}\\
\hline
00 & 1.88\% & 1.88\% & 12 & 4.86\% & 4.86\% \\
01 & 2.40\% & 2.40\% & 13 & 5.70\% & 5.70\% \\
02 & 2.92\% & 2.92\% & 14 & 4.49\% & 4.49\% \\
03 & 3.56\% & 3.56\% & 15 & 4.46\% & 4.46\% \\
04 & 3.73\% & 3.73\% & 16 & 3.94\% & 3.94\% \\
05 & 3.79\% & 3.79\% & 17 & 4.31\% & 4.31\% \\
06 & 4.80\% & 4.80\% & 18 & 3.53\% & 3.53\% \\
07 & 5.79\% & 5.79\% & 19 & 3.10\% & 3.10\% \\
08 & 6.54\% & 6.54\% & 20 & 3.36\% & 3.36\% \\
09 & 6.14\% & 6.14\% & 21 & 3.07\% & 3.07\% \\
10 & 6.69\% & 6.69\% & 22 & 2.49\% & 2.49\% \\
11 & 5.76\% & 5.76\% & 23 & 2.69\% & 2.69\% \\

\bottomrule
\end{tabular}}
\vspace{-1mm}
\end{table}

\begin{table}{}%
\fontsize{12pt}{12pt}\selectfont
\caption{Minutes in an Hour.}
\vspace{-0.07in}
\label{tab:Minutes in an Hour}
\centering
\scalebox{0.6}{
\begin{tabular}{clclclclc}
  \toprule
   \multicolumn{1}{c}{Minute} &  \multicolumn{1}{c}{Analysts} & \multicolumn{1}{c}{GPT-4} & \multicolumn{1}{c}{Minute} & \multicolumn{1}{c}{Analysts} & \multicolumn{1}{c}{GPT-4}\\
\hline
0 & 1.13\% & 1.12\% & 30 & 2.08\% & 2.08\% \\
1 & 1.91\% & 1.91\% & 31 & 2.08\% & 2.08\% \\
2 & 1.77\% & 1.76\% & 32 & 1.45\% & 1.44\% \\
3 & 1.56\% & 1.56\% & 33 & 1.59\% & 1.59\% \\
4 & 1.62\% & 1.62\% & 34 & 1.68\% & 1.67\% \\
5 & 2.08\% & 2.08\% & 35 & 1.51\% & 1.50\% \\
6 & 1.45\% & 1.44\% & 36 & 1.74\% & 1.73\% \\
7 & 1.71\% & 1.70\% & 37 & 1.48\% & 1.47\% \\
8 & 1.82\% & 1.82\% & 38 & 1.24\% & 1.24\% \\
9 & 1.42\% & 1.41\% & 39 & 1.91\% & 1.91\% \\
10 & 1.88\% & 1.88\% & 40 & 1.68\% & 1.67\% \\
11 & 1.65\% & 1.64\% & 41 & 1.45\% & 1.44\% \\
12 & 1.56\% & 1.56\% & 42 & 1.59\% & 1.59\% \\
13 & 1.62\% & 1.62\% & 43 & 1.71\% & 1.70\% \\
14 & 1.48\% & 1.47\% & 44 & 1.85\% & 1.85\% \\
15 & 2.00\% & 1.99\% & 45 & 1.79\% & 1.79\% \\
16 & 1.27\% & 1.27\% & 46 & 1.82\% & 1.82\% \\
17 & 1.45\% & 1.44\% & 47 & 1.71\% & 1.70\% \\
18 & 1.53\% & 1.53\% & 48 & 1.42\% & 1.41\% \\
19 & 1.27\% & 1.27\% & 49 & 2.00\% & 1.99\% \\
20 & 2.05\% & 2.05\% & 50 & 1.71\% & 1.71\% \\
21 & 1.53\% & 1.53\% & 51 & 1.45\% & 1.44\% \\
22 & 2.08\% & 2.08\% & 52 & 1.48\% & 1.47\% \\
23 & 1.65\% & 1.64\% & 53 & 1.07\% & 1.07\% \\
24 & 1.68\% & 1.67\% & 54 & 2.05\% & 2.05\% \\
25 & 1.33\% & 1.33\% & 55 & 1.45\% & 1.44\% \\
26 & 1.42\% & 1.42\% & 56 & 2.05\% & 2.05\% \\
27 & 1.88\% & 1.88\% & 57 & 1.82\% & 1.82\% \\
28 & 2.14\% & 2.14\% & 58 & 1.39\% & 1.38\% \\
29 & 1.85\% & 1.85\% & 59 & 1.94\% & 1.93\% \\

\bottomrule
\end{tabular}}
\vspace{-1mm}
\end{table}

\vspace{-0.09in}
\subsection{Results: Domain-Specific Analyses} 
\label{sec:results:domain-specific analysis}

The understanding of domain-specific analytics is and has continuously been investigated by the research community \cite{li2023chatgpt}, \cite{goetz2006domain}, \cite{agarwal2023financial}.
Goetz et al. \cite{goetz2006domain} examines the domain specificity of emotions with a focus on experiences of enjoyment, anxiety, along with boredom in the academic domains of English, Latin, German and Mathematics. Goetz et al. \cite{goetz2006domain} notice that emotions are significantly domain-specific  than  students’ grades.
Agarwal \cite{agarwal2023financial}, remarks that generally, embeddings trained on
general data do not perform well on domain-specific tasks. Agarwal \cite{agarwal2023financial}, further notes that the effectiveness of term representation methods, e.g., word embeddings greatly depends on two factors: (1) the domain-specific task and (2) the kind of data used to train the embeddings.  
This study, in its quest for knowledge, examines GPT-4 by exploring its ability at conducting domain-specific analysis in the realm of NLP. 

\subsubsection{Verbs and Nouns GPT-4}
The study observes that GPT-4 is not certain on how to output the maximum, minimum and mean values in relation to the verb, as well as noun counts for both the subject lines and body messages of the phishing emails. 
GPT-4 attributes the reason for the uncertainty to be the lack of specific libraries i.e., \textit{TexBlob} a Python (2 and 3) \cite{loria2018textblob}, as well as \textit{nltk} a platform for working in computational linguistics domain using Python \cite{loper2002nltk}.
In trying to attain comparison graphs similar to Figures \ref {fig:Word_verb_noun_count_subject_lines} and  \ref{fig:Word_verb_noun_countbBody_messages}. GPT-4, calls for the installation of the necessary libraries and in particular \textit{nltk} \cite{loper2002nltk}, as well as \textit{matplotlib}, a comprehensive plotting library for creating static, animated, and interactive visualizations in Python and its numerical mathematics extension NumPy \cite{hunter2007matplotlib}.

\subsubsection{Emotion Analysis}
\begin{figure}[t!]
    \centering
    \includegraphics[scale=0.75]{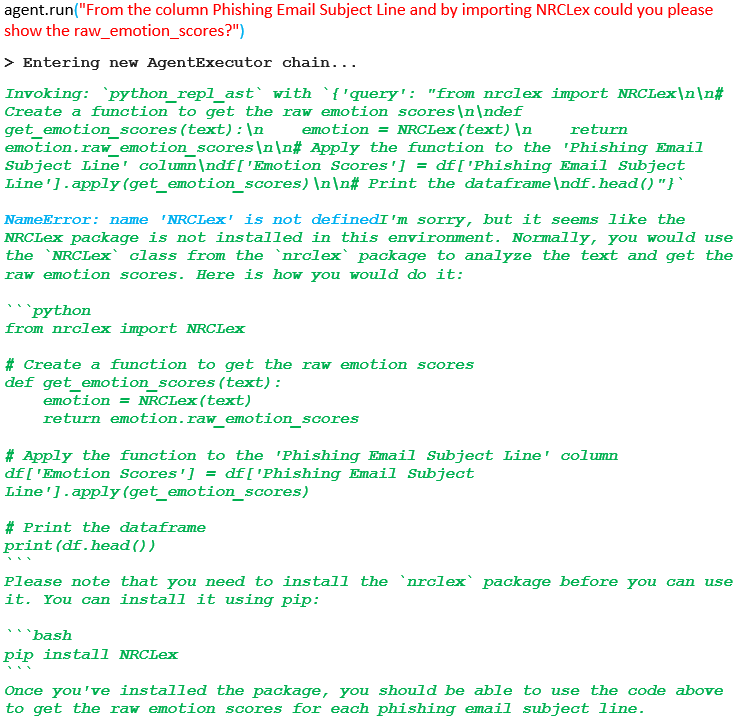}
    \caption{GPT-4 NRCLex Emotional Analysis.}
    \label{fig:GPT-4 NRCLex Analysis}
 \vspace*{-0.17in}
\end{figure}

Trust at 34.69\% and at 29.97\% is observed to be the top emotion by the analysts from the subject lines and body messages of the phishing emails respectively. Table \ref{tab:Emotional Affect Frequencies}, The Analysts Emotional Affect Frequencies, further illustrates that phishing emails employ different forms of emotional manipulation tactics i.e., fear, anger, anticipation, surprise, sadness, disgust, as well as joy to deceive the unsuspecting victims. 
This study additionally explores GPT-4 for emotional analysis of the phishing emails using NRCLex, and Figure \ref{fig:GPT-4 NRCLex Analysis}, extracted from GPT-4 reports that GPT-4 suggests that it seems like the NRCLex package is not installed in its environment. However, it (GPT-4) recommends progressive steps to help with the tasks of phishing emails emotional analysis using NRCLex.   

\begin{table}{}%
\fontsize{12pt}{12pt}\selectfont
\caption{The Analysts Emotional Affect Frequencies.}
\vspace{-0.07in}
\label{tab:Emotional Affect Frequencies}
\centering
\scalebox{0.6}{
\begin{tabular}{clclclclc}
  \toprule
   \multicolumn{1}{c}{Class} &  \multicolumn{1}{c}{Subject} & \multicolumn{1}{c}{Body} &  \multicolumn{1}{c}{Class} & \multicolumn{1}{c}{Subject} &  \multicolumn{1}{c}{Body}\\
\hline
Fear & 5.33\% & 5.22\% & Positive & 21.79\% & 27.15\%\\
Anger & 3.28\% & 3.62\% & Negative & 13.29\% & 12.04\%\\
Anticipation & 11.12\% & 9.21\% & Sadness & 6.16\% & 5.50\%\\
Trust & 34.69\% & 29.97\% & Disgust & 0.35\% & 1.96\% \\
Surprise & 2.68\% & 1.89\% & Joy & 1.31\% & 3.44\%\\
\bottomrule
\end{tabular}}
\vspace{-1mm}
\end{table}

\vspace{-0.09in}
\section{Conclusion and Future Work} 
\label{sec:conclusion}

Large Language Models (LLMs) are reshaping society. While studies have explored their ability to generate human-like responses, their potential as statistical personal assistants requires further investigation. This study examines LLMs' performance in statistically driven descriptive evaluations, serving as personal assistants to users with limited domain knowledge. Focusing on NLP-based investigations of cybersecurity texts, the paper compares results generated by analysts to those from interactions with Langchain and GPT-4. Findings show that GPT-4 performs well in numerical reasoning and feature engineering tasks, but exhibits some limitations on tasks such as: (1) phishing emails emotional analysis using specified emotion analytics domain-specific knowledge package, (2) correlation matrices generation, in addition to (3) polarity classification. The study concludes by suggesting the need for replication, as well as extension of the experiments into various realms to fully assess the capabilities of LLMs as personal assistants and or for domain-specific tasks.

\vspace{-0.09in}
\section*{Acknowledgment}
This research was supported by the U.S. National Science Foundation (Awards\#: 2319802 and 2319803). Opinions, findings, and conclusions are those of the authors and do not necessarily reflect the views of the NSF.

\bibliographystyle{IEEEtran}
\vspace{-0.09in} \bibliography{mybibdada24}

\end{document}